# Hyades star cluster and the New comets

M. D. Sizova[1], E. S. Postnikova[1], A. P. Demidov[2], N. V. Chupina[1],
S. V. Vereshchagin[1]

[1]Institute of Astronomy, Russian Academy of Sciences, Pyatnitskaya str., 48, 119017 Moscow, Russia
[2]Central Aerological Observatory, Pervomayskaya str., 3, 141700 Dolgoprudny, Moscow region, Russia
sizova@inasan.ru es_p@list.ru the-admax@ya.ru chupina@inasan.ru svvs@ya.ru

**Abstract.** We examined the influence of the Hyades star cluster on the possibility of the appearance of long-period comets in the Solar system. It is known that the Hyades cluster is extended along the spatial orbit on tens of parsecs. To our estimations, 0.85 million years ago, there was a close approach of the cluster to the Sun of 24.8 pc. The approach of one of the cluster stars to the Sun at the minimally known distance of about 6.9 pc was 1.6 million years ago. The main part of the cluster was close to the Sun from 1 to 2 million years ago. Such proximity is not essential for the impact on the dynamics of small bodies in the external part of the Oort cloud, although the view may change after additional study of the cluster structure. Possible orbits perihelion displacements of the small bodies of the outer part of the Oort cloud make some of them in observable comets region.

**Introduction.** The data obtained by Gaia mission allows us to study previously inaccessible details of the structure of stellar systems. Thus, it was found that the Hyades star cluster is extended along the spatial orbit by tens of parsecs (Röser et al. 2019, Meingast and Alves 2019). As it turned out, its spatial size is so large that in the past its stars could come closer to the Solar system. The approaches of stars with the Sun affect the parameters of the orbits of small bodies that populate the periphery of the planet systems. Many stars - candidates for such proximity rapprochement both in the past and in the future - have been discovered by (Torres et al 2019). May occur 1) moving the Oort comets to orbits closest to the Sun 2) the loss of small bodies by the outer parts of the Solar system and the system of planets of other stars. In the first case may appear the observed comets, in the second one - interstellar small bodies.
Does star clusters participate in these processes? Having a mass hundred times greater than the mass of an individual star, the cluster can cause orbits parameters changes of small bodies in the outer parts of the Oort cloud and, accordingly, the appearance of many long-period comets.

**The work structure.** The data we used. Spatial-kinematic structure of Hyades star cluster. Hyades orbit in the galactic disk. Rapprochement with the Solar system. LB-diagram, Hyades orbit on the LB-diagram. The relation between Hyades and the appearance of the New comets. Conclusions.

**The data we used.** In order to evaluate the motion of not only the cluster but also of individual stars, including those far from its center, we required a list of reliably selected stars with measured spatial coordinates and velocities. We took such a catalog, based on Gaia DR2 data, in

Meingast and Alves 2019. As comets list Table Matese and Whitmire 2011 is used (Appendix B: Orbital elements of all 17th Catalog class 1A comets with x <100. All angles are in degrees, from the lists 102 comets with a> 104 AU).

**Spatial-kinematic structure of Hyades star cluster.** Fig. 1 shows the position of the Hyades stars in the galactic Cartesian coordinate system near 2 Myr ago. Arrows indicate the projections of the spatial velocity vectors of individual stars. The Sun is in the center.

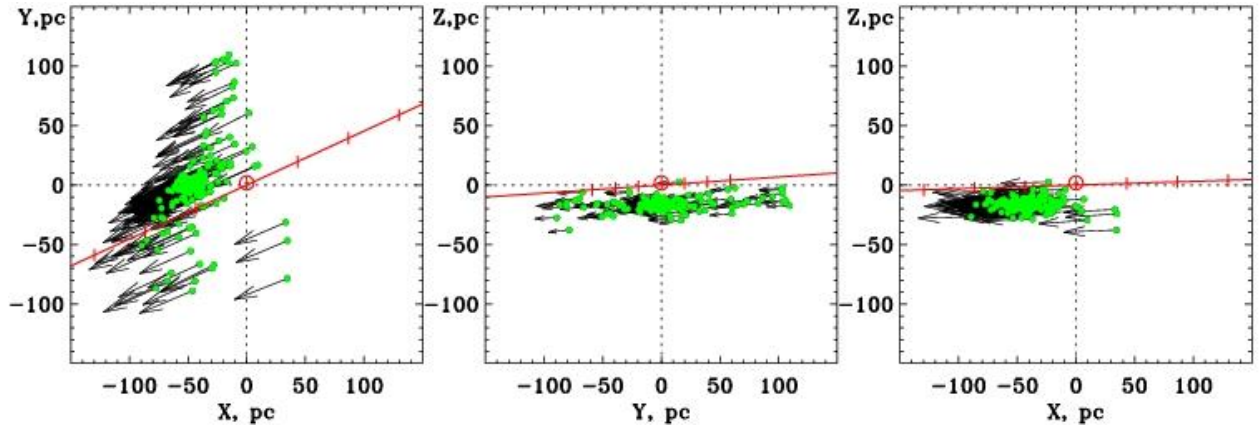

Fig. 1. Hyades in the galactic coordinate system. The X-axis is directed towards to the Galaxy center (l = 0°, b = 0°), the Y-axis is in the direction of the Galaxy rotation (l = 90°, b = 0°), the Z-axis is toward to the Galaxy north pole (b = 90°). The list of stars - members of the cluster is taken from Meingast and Alves 2019. The green dots indicate the positions of the stars, arrows - the projections of their spatial velocity vectors. The red line is the movement of the Sun, strokes show the intervals of 1 million years.

**Hyades orbit in the galactic disk. Rapprochement with the solar system.** The integration of the cluster center and Sun motions showed they had a close approach at a distance of approximately 24.8 pc from 0.8 to 2 million years ago. Our calculations showed that individual Hyades stars approached to the Sun at a distance of less than 7 pc. The obtained distances were calculated separately for each star in the cluster according to the results of integration in the Milky Way Galaxy potential using Galpy, Bovy 2015.

**Hyades orbit on the LB diagram.** It is known, the position of the entry points Oort comets into the Solar system align with their position in the sky. Thus, it is possible to find-stars that served as sources of perturbation of the Oort cloud due to their close passage to the solar system. Of greatest interest are the new comets (New comets - these are Oort comets, approaching the Sun for the first time).

Fig. 2 shows the positions of the new comets aphelions (the orbital inclination *i* is close to 103°) and the positions calculated for the center of Hyades 3 million years ago (black dashed line), and also some noteworthy points. In the time interval from 1 to 2 million years ago, the position of Hyades overlapped with a group of comets in the red arch.

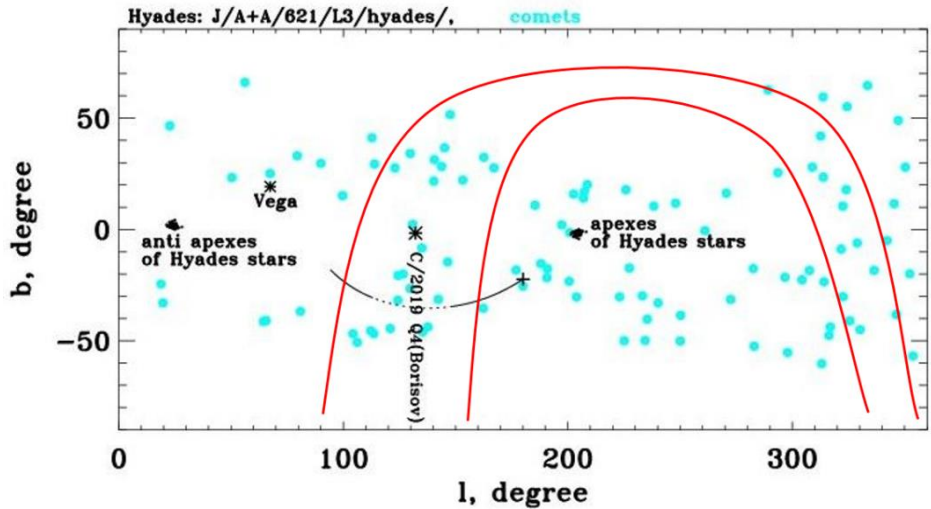

Fig. 2. LB-diagram. The track (in black dotted lines) indicates the displacement of the Hyades cluster center 3 million years ago. Areas of the apex and anti-apex of Hyades are signed. The red lines show the position of a new comets region with a binding energy of 30 <x <60 and orbital inclinations close to 103° (the appearance of these comets could have caused by hypothetical massive Sun satellite of located inside the Oort cloud). Comet positions according to the Matese table, Whitmire 2011 (Appendix B: Orbital elements of all 17th Catalog class 1A comets with x <100. All angles are in degrees, from the lists 102 comets with A> 104 AU).

**The relation between Hyades-Sun close approach and the appearance of the New comets.**
To evaluate the separation possibility of comets from the outer part of the Oort cloud by a star cluster passing by the Solar system let $M_{Sun}$ be the mass of the Sun, r be the distance of the comet from the Sun, approximately 0.5 pc - 1.0 pc. The velocity of comet separation from the Solar System (second space velocity): $v^2_{escape} = 2\ G\ M_{Sun}/r$.
In our case, it equal $v_{escape} = 133$ m / s.

Let's consider the value of the change of the small body velocity at the outer boundary of the Oort cloud caused by a random close passage of a star or star cluster $\Delta V_\perp$, Torres et al 2019:
$\Delta V_\perp = 0.1 \cdot M_{cl}/r^2_{Sun}$

The increment of the small body velocity-$\Delta V_\perp$ at a given distance of the small body from the Sun (we used 0.5 pc) depends on the cluster mass and the approach distance Sun-Hyades $r_{Sun}$. The Hyades cluster, whose mass is approximately 400 $M_{Sun}$, approaching the Sun by 24.8 pc, leads in the small bodies velocities located at the outer boundary of the Oort cloud to an increase by approximately 0.1 m / s. The escape velocity here is 133 m / s.

**Conclusions.** Could such a rapprochement between the Sun and the Hyades cluster affect the orbit parameter change of many small bodies from the Oort cloud? Only under certain conditions. We still cannot say about their implementation with full certainty. It is necessary to study the cluster structure both for adding weak stars and more accurate space location in space. It is known that open star clusters with time can extended along orbits over distances of hundreds of parsecs. If the following conditions exist: more than a thousand stars are in the cluster; its center is located at a distance of 24.8 pc from the Sun; the comet is distant from the

Sun no further than 0.5 pc; velocity $\Delta V_\perp = 0.1$ m/s, then the collective influence of the cluster stars can increase the velocity increment by at least 2.5 times.

When approaching a single star to the Sun is 6.9 pc, the velocity increment is equal $\Delta V_\perp = 0.002$ m/s. It is not enough for the comet to leave the Oort cloud. If we take into account the stars stream formed by cluster tidal tails can contain up to 1000 stars, then the total effect will be much larger, so it needs to be calculated. In reality scatter of the tidal tail can be up to 600 pc; the detection of cluster member stars in this region can lead to an increase of $M_{cl}$, that accordingly will lead to increase $\Delta V_\perp$.

Hyades passed "under" the Solar system in the galactic disk, which could affect the appearance of comets with high inclinations of the orbits relative to the galactic plane (i = 103°). About a completely different reason - the existence of a hypothetical satellite of the Sun in the Oort cloud, was discussed above.

We note that the velocities dispersion of the Hyades stars, equal to 0.4 km / s, will lead to stars relocation within the cluster. Such displace for 2 million years can be up to 1 pc for one star. This is the minimum accuracy estimate for the distance values we calculated. The search for weak cluster members, taking into account the mass of binaries and the collective influence of the stars of the Hyades cluster will increase the accuracy of this estimate.